\documentclass[aps,prl,reprint]{revtex4-1}
\usepackage[utf8]{inputenc}

\usepackage{graphicx}
\usepackage{dcolumn}
\usepackage{bm}
\usepackage{epstopdf}
\usepackage{color}
\usepackage{amsmath, amssymb}
\usepackage[hidelinks]{hyperref}

\begin{document}

\title{Quantitative Measurement of Density in a Shear Band of Metallic Glass Monitored along its Propagation Direction}

\author{Vitalij Schmidt}
\email{vitalij.schmidt@uni-muenster.de}
\affiliation{Institut für Materialphysik, Westfälische Wilhelms-Universität  Münster, Wilhelm-Klemm-Str. 10, D-48149 Münster, Germany}
\author{Harald Rösner}
\affiliation{Institut für Materialphysik, Westfälische Wilhelms-Universität  Münster, Wilhelm-Klemm-Str. 10, D-48149 Münster, Germany}
\author{Paul M. Voyles}
\affiliation{Materials Science and Engineering, University of Wisconsin-Madison, 1509 University Ave, Madison, Wisconsin 53706, USA}
\author{Martin Peterlechner}
\affiliation{Institut für Materialphysik, Westfälische Wilhelms-Universität  Münster, Wilhelm-Klemm-Str. 10, D-48149 Münster, Germany}
\author{Gerhard Wilde}
\affiliation{Institut für Materialphysik, Westfälische Wilhelms-Universität  Münster, Wilhelm-Klemm-Str. 10, D-48149 Münster, Germany}

\begin{abstract}
Quantitative density measurements from electron scattering show that shear bands in deformed Al$_{88}$Y$_7$Fe$_5$ metallic glass exhibit alternating high and low density regions, ranging from -9~\% to +6~\% relative to the undeformed matrix. Small deflections of the shear band from the main propagation direction coincide with switches in density from higher to lower than the matrix and vice versa, indicating that faster and slower motion (stick slip) occurs during the propagation. Nanobeam diffraction analyses provide clear evidence that the density changes are accompanied by structural changes suggesting that shear alters the packing of tightly bound short- or medium-range atomic clusters. This bears a striking resemblance to the packing behavior in granular shear bands formed upon deformation of granular media.
 
\begin{description}
 \item[PACS numbers]
 \verb+ 61.05.J-, 61.43.Dq, 83.50.-v, 83.80.Ab +
\end{description}
\end{abstract}


\maketitle

Metallic glasses (MGs) offer unique properties such as high strength, extended elasticity, high wear and corrosion resistance, or excellent soft magnetic behavior~\cite{Ashby2006}. However, the limited ductility and especially the immediate catastrophic failure in tension once the elastic limit is reached are major obstacles to applications as structural materials~\cite{Ashby2006}. This behavior has led to substantial effort towards understanding and improving the accommodation of plastic deformation in MGs, including the design of composites consisting of ductile crystalline phases in a bulk MG matrix~\cite{Hays2000,Szuecs2001,Donohue2007,Eckert2007,Hofmann2008}. Monolithic glasses with higher Poisson’s ratios near the ideal value of 0.5~\cite{Schroers2004,Lewandowski2005,Demetriou2011} tend to have improved compressive and bending plasticity, and are reported to show a fine dispersion of shear bands~\cite{Wang2011}. Well below the glass transition temperature, plastic flow in metallic glasses is restricted to narrow regions called shear bands~\cite{Spaepen1977, Argon1979, Greer2013}, widely believed to be dilated zones of increased free volume caused by shear localization enabling shear softening~\cite{Donovan1981, Klaumunzer2011, Pan2011}. Thus, the density of or free volume inside shear bands is often treated as key to understanding plasticity in MGs. Quite a large range of density or free volume changes have been observed, based on various experiments~\cite{Nagel1998,Wilde2000,Chen2009,Zhang2009,Dmowski2010,Lechner2010,Klaumunzer2011,Pan2011,Shao2013}. However, most experimental methods, including positron annihilation spectroscopy and calorimetry, measure the free volume integrated over the entire sample to be typically from +1~\% to +3~\%. High resolution transmission electron microscopy has shown the presence of nanovoids inside shear bands, believed to form from the coalescence of atomic-sized free volumes after deformation~\cite{Li2002}, but so far has not been used to measure material density either on average or locally within the shear band.

Previously the local density within shear bands in Al$_{88}$Y$_7$Fe$_5$ has been probed at selected positions and for different shear bands using quantitative HAADF-STEM on deformed and undeformed metallic glass regions (see Supplemental Material for details). These experiments showed both positive and negative density changes for shear bands with respect to the undeformed glass~\cite{Rosner2014}. Minor compositional changes relative to the matrix were observed in the shear band segments. While compositional changes contributed mainly to the positive density variations in the bright shear band segments, the change in free volume was the dominant effect for the dark shear band segments.

Here we show, by combining several quantitative electron scattering signals, that shear bands in Al$_{88}$Y$_7$Fe$_5$ metallic glass have segments of both decreased and increased density that alternate along the propagation direction. The changes in density are correlated to small deflections in the propagation direction and changes in medium-range structural order and chemical composition. Similar behavior has been reported for granular media, but not for amorphous solids. One important implication is that individual shear bands apparently propagate via local, segment-wise stick slip that might originate from their complex topology. These results are important to the physics of deformation in metallic glasses, and suggest connections to the physics of granular materials and jammed systems.

Ingots of the target composition Al$_{88}$Y$_7$Fe$_5$ were prepared by arc melting pure Al (99.999~\%), Fe (99.99~\%) and Y (99.9~\%). Fully amorphous ribbons with a thickness of about 40 $\mu$m were produced by melt spinning using a tangential wheel (Cu) speed of 47~ms$^{-1}$. The melt-spun ribbons were cold-rolled to a thickness reduction of 23~\% and subsequently prepared for transmission electron microscopy (TEM) by twin-jet electropolishing using HNO$_3$:CH$_3$OH in a ratio 1:2 at 253~K. The TEM study was performed with a FEI Titan 80-300 (probe) aberration-corrected transmission electron microscope operated at 300~kV in the scanning transmission electron microscopy (STEM) mode. The following conditions were used during the experiments: a probe current of 15~pA, a collection semiangle of 51.3~mrad for the high-angle annular dark-field (HAADF) detector (130~mm camera length), a convergence semiangle $\alpha$ of 0.6~mrad, a nominal spot size of 2~nm, and a step size of 2~nm. For electron energy loss spectroscopy (EELS) we used a collection semiangle $\beta$ of 5.73~mrad, an entrance aperture of 2.5~mm, an energy dispersion of 0.1~eV/channel and an acquisition time of 0.025~s. The fluctuation electron microscopy (FEM) data were obtained from energy-filtered nanobeam diffraction using a camera length of 510~mm.

\begin{figure}
 \includegraphics[width=8.6cm]{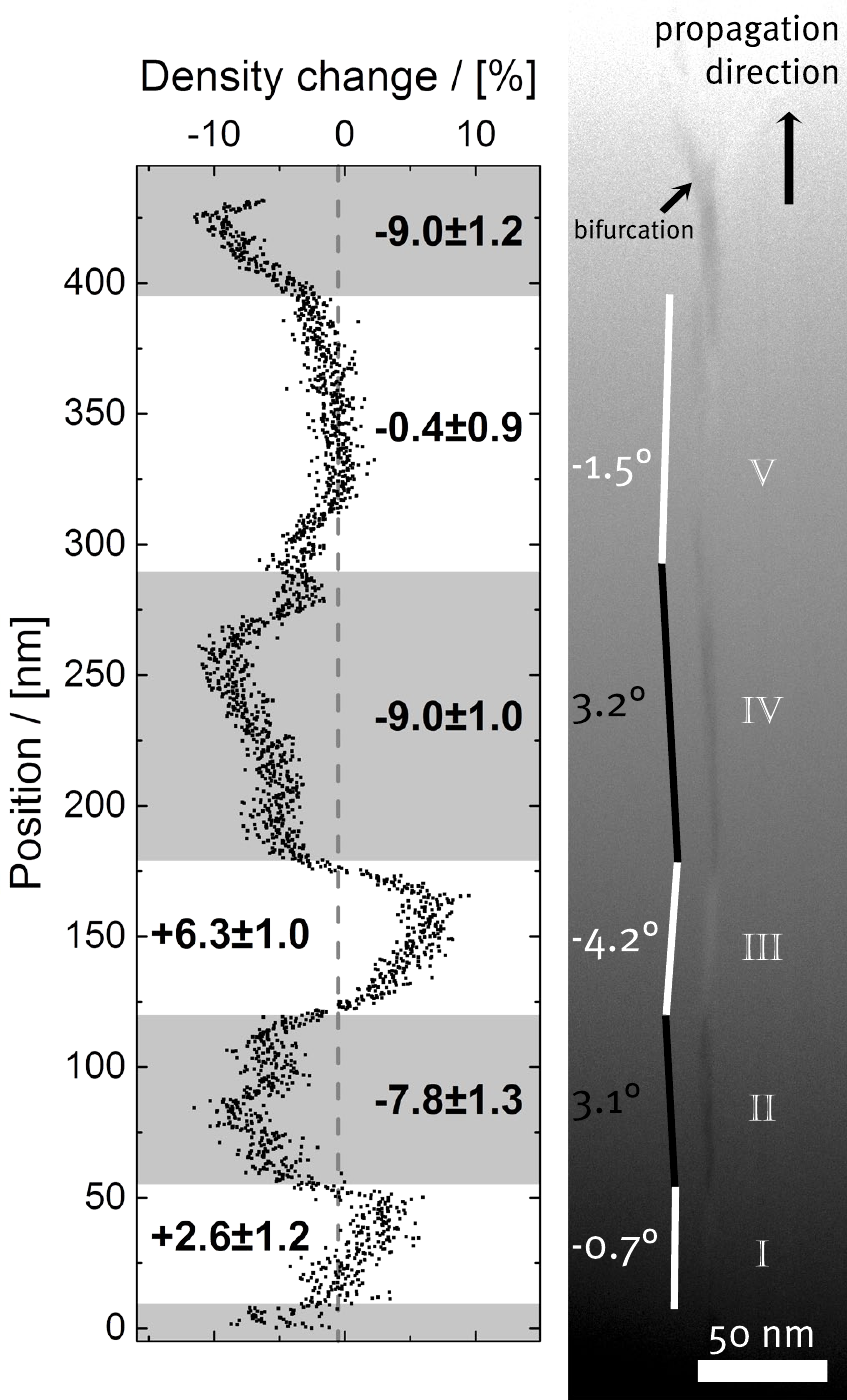}
 \caption{\textbf{Density variations in a shear band.} Right: A HAADF-STEM image showing a shear band in cold-rolled Al$_{88}$Y$_7$Fe$_5$ propagating from bottom to top with several contrast changes. The individual segments are numbered ($I-V$). The deflection angles are indicated for the segments. Left: Quantified density variations of the shear band along the propagating direction with respect to the undeformed matrix. Bright and dark parts of the shear band are indicated by the white and gray background, respectively.}
 \label{fig:Fig_1}
\end{figure}

HAADF-STEM images from deformed Al$_{88}$Y$_7$Fe$_5$, such as shown in Fig.~\ref{fig:Fig_1}, frequently showed shear bands with contrast changes from bright to dark and vice versa along their propagation direction. Each contrast reversal is accompanied by a slight deflection~\cite{Rosner2014}, always within $\pm 5^\circ$ of the main propagation direction of the shear band. The propagation direction is determined from the bifurcation. The angular deflection range matches the angle between the shear bands and the direction of the applied shear stress reported in literature~\cite{Greer2013}.

\begin{figure}
 \includegraphics[width=8.6cm]{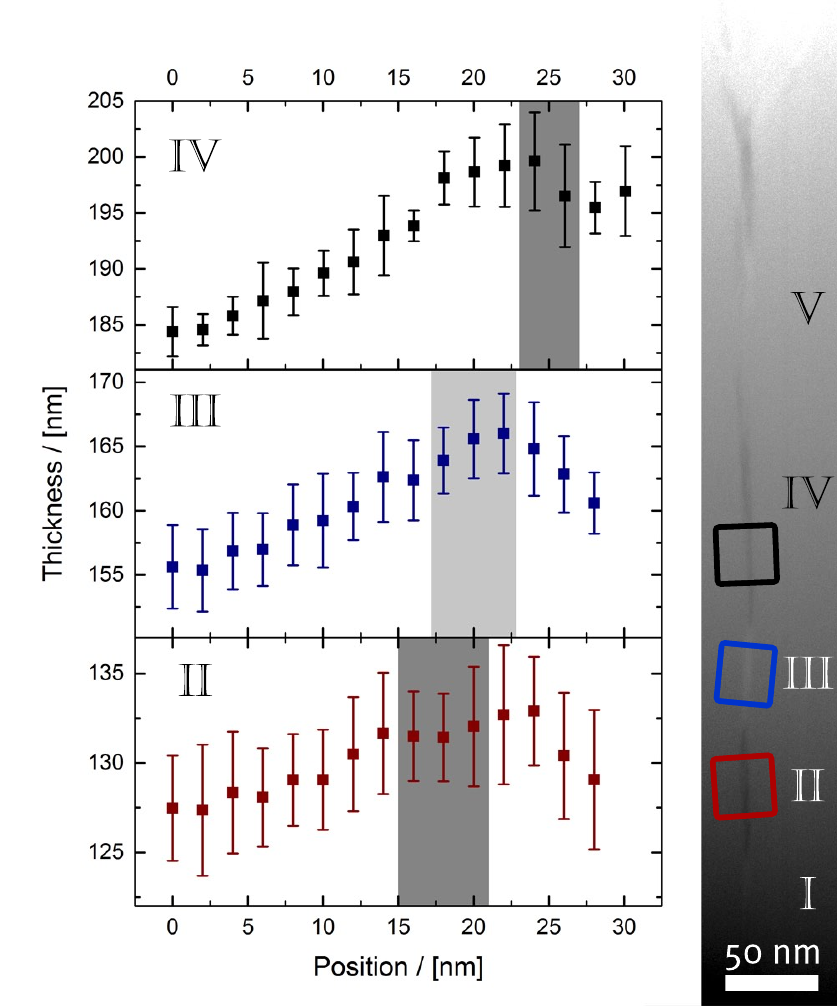}
 \caption{\textbf{Foil thickness calculations.} Profiles of the averaged foil thicknesses for the boxed regions of the HAADF-STEM image displaying different shear band parts (see segments $II$, $III$, and $IV$ in Fig.~\ref{fig:Fig_1}). The gray shaded areas indicate the position of the shear band.}
 \label{fig:Fig_2}
\end{figure}

Figure~\ref{fig:Fig_2} shows the foil thickness profiles from part of the sample area shown in Fig.~\ref{fig:Fig_1} determined from low-loss EELS. There is a linear increase in foil thickness along the propagation direction of the shear band which is related to the wedge shape of the TEM specimen. There is also a slight thickness variation between the two sides of the shear band, probably due to the shear offset. Thickness profiles across the shear band show that there is no preferential thinning of the shear band by TEM sample preparation.

The density variation $\Delta \rho$ of each bright or dark segment was determined from the HAADF intensity and EELS thickness, as described in detail in the Supplemental Material. The left side of Fig.~\ref{fig:Fig_1} shows the local density inside the shear band along its propagation direction. Repetitive density changes are observed. The density differences between the shear band and undeformed matrix range from $(-9.0 \pm 1.0)~\%$ to $(+ 6.3 \pm 1.0)~\%$. The matrix region outside the shear bands does not show any impact of the deformation or of preexisting density fluctuations. The integrated density change along this specific shear band is $\Delta \rho = \left( -2.9\pm 4.3 \right)~\%$. No characteristic segment length was found over examination of many shear bands. However, the dark low density segments dominate the shear bands in general, which results in a negative value for the integrated density change. The mean density is in good agreement with previous reports of shear band density from macroscopic experiments~\cite{Lechner2010,Klaumunzer2011,Shao2013}. The width of the shear band was measured on the clearest segments ($II$, $III$, and $IV$) by fitting Gaussian peaks to the data in Fig.~\ref{fig:Fig_1}. The FWHM of $w_{II} = (3.72 \pm 0.21)$~nm, $w_{III} = (4.26 \pm 0.16)$~nm and $w_{IV} = (4.24 \pm 0.15)$~nm. Segment $V$ is wider by inspection, so the consistency of the width in the other segments shows that they image the shear band near edge-on. When the shear band is inclined, the projected density difference along the electron beam direction is suppressed due to the overlap of shear band and matrix.

At this point it might be appropriate to question whether the shear band structures analyzed resemble the structures formed during the shear band propagation. Since no crystal formation was found for the denser regions of the shear band, where crystallization may be expected to be facilitated~\cite{Wilde2003}, it seems safe to assume that the shear bands analyzed in this work do closely resemble the structures present during deformation.

\begin{figure}
 \includegraphics[width=8.6cm]{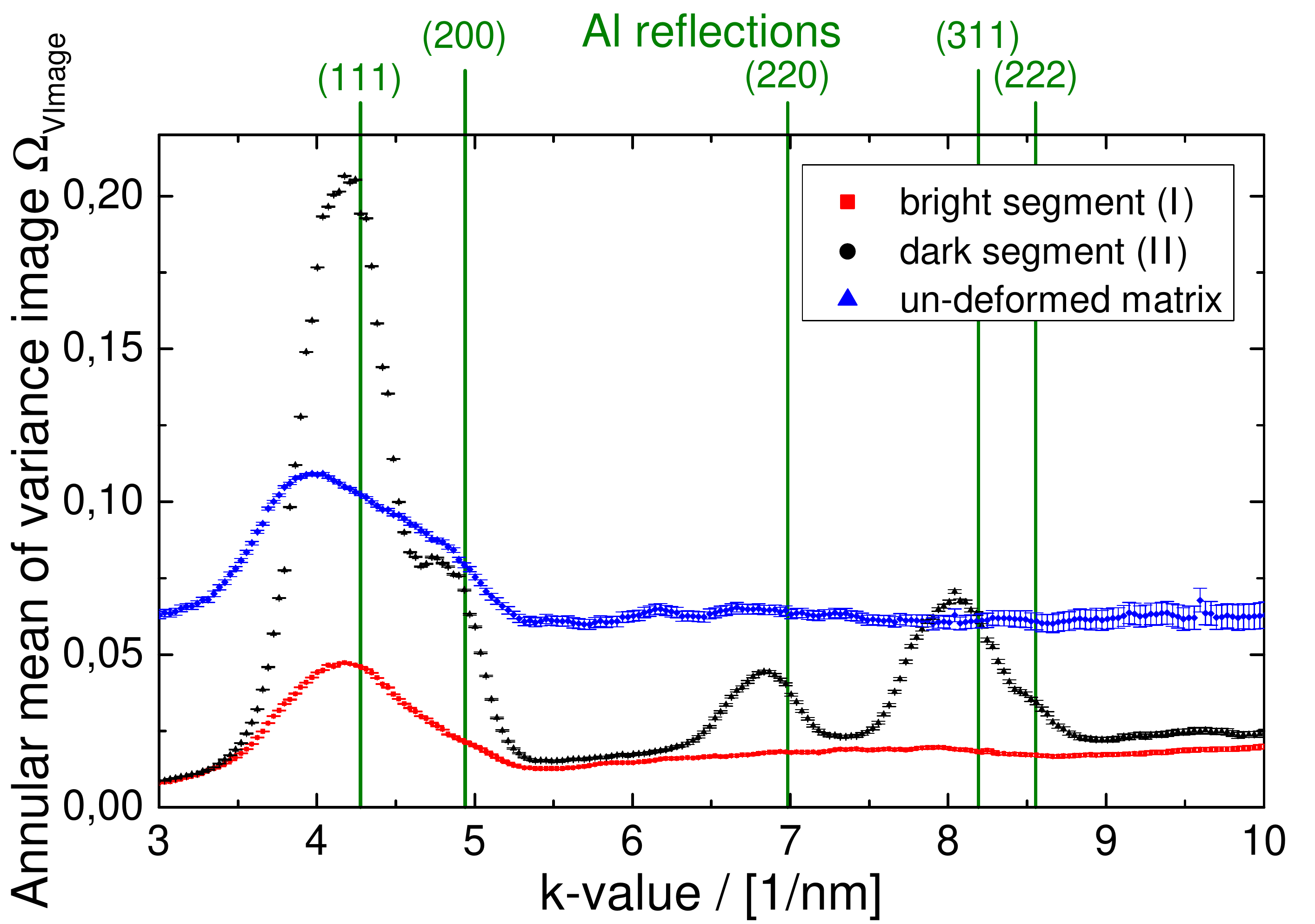}
 \caption{\textbf{Fluctuation Electron Microscopy.} Annular mean of variance image $\Omega_\textrm{VImage}$ of different NBDP ensembles (undeformed matrix, bright/dark parts of the shear band corresponding to the segments $I$ and $II$).}
 \label{fig:Fig_3}
\end{figure}

FEM provides information about the nanoscale order of the metallic glass from systematic coherent electron nanodiffraction (see Supplemental Material)~\cite{Daulton2010,Yi2012}. Nanoscale order leads to variability in nanodiffraction into different directions in reciprocal space, which is captured quantitatively in the annular mean of the variance image $\Omega_\textrm{VImage}(k)$~\cite{Daulton2010}. Figure~\ref{fig:Fig_3} shows FEM data from a bright high-density segment (segment $I$), a dark low-density shear band segment (segment $II$), and the undeformed matrix about 20~nm away from the dark shear band segment. The results for the lower density, dark segment are similar to our previous report~\cite{Rosner2014}, but we have obtained better statistical quality data for the higher density, bright segment here. After accounting for differences in sample thickness, the variance signal for the matrix is consistent with previous FEM studies of the same glass composition~\cite{Yi2012}, which data showed that undeformed Al$_{88}$Y$_7$Fe$_5$ contains nanoscale ordered regions with an internal structure similar to fcc Al embedded in a more disordered matrix. The primary peak in $\Omega_\textrm{VImage}(k)$ sits near the Al $\left\lbrace111\right\rbrace$ reflection, there is a shoulder at higher $k$ consistent with Al $\left\lbrace200\right\rbrace$, and there are no peaks at higher $k$. $\Omega_\textrm{VImage}(k)$ for the bright, higher density shear band segment is similar to the matrix, but the $\left\lbrace200\right\rbrace$ shoulder is suppressed. This indicates a similar structure of Al-like regions embedded in a matrix, but with reduced size or internal structural order for the Al-like regions~\cite{Yi2012,Rosner2014}. $\Omega_\textrm{VImage}(k)$ for the dark, low density shear band segment indicates strikingly high structural order. In addition to clear $\left\lbrace 111\right\rbrace$ and $\left\lbrace200\right\rbrace$ peaks, there are peaks at the $\left\lbrace311\right\rbrace$ and $\left\lbrace220\right\rbrace$ Al reflections as well~\cite{Rosner2014}. However, all peaks are shifted towards lower $k$ values. The reason for this has been shown to be hydrostatic tensile strain originating from the surrounding amorphous material in which the Al-like regions or clusters are embedded~\cite{Yi2012}. The increase in $\Omega_\textrm{VImage}(k)$ shows that the Al-like regions are larger than in the matrix, and the persistence of peaks to high $k$ shows that they have a higher degree of crystalline structural order~\cite{Yi2012}. Strong Al-like order is consistent with previous observations of Al crystallite nucleation in shear bands~\cite{Wilde2003}.
Here we emphasize that we have a mixture composed of Al-rich cluster/crystallites embedded in amorphous material present in the dark shear band segments rather than fully crystallized regions since the latter would produce enormous peaks in the fluctuation signal which we do not observe. Moreover, a former study showed that fully crystallized Al$_{88}$Y$_7$Fe$_5$ metallic glass is composed of three crystalline phases, that is, fcc Al, Al$_3$Y (D$0_{19}$ structure) and a ternary intermetallic Al$_7$Fe$_5$Y phase~\cite{Boucharat2005}. Since the diffraction peaks of the FEM study closely match fcc Al with the other two crystalline phases missing, a complete crystallization of the dark shear band segment can be excluded here. We estimate the size of the ordered Al-like regions to be $\leq$2~nm from the similarity of adjacent nanodiffraction patterns~\cite{Hirata2014}. Since we used a 2~nm diameter probe for the experiments, 2~nm is an upper bound and the real ordered regions may be even smaller. Moreover, for 3d transition metals such sizes are commonly associated with clusters rather than with the crystalline state.

Larger density decreases of 9~\% or more as reported here must be accommodated by some local chemical changes, not just an increase in mean atomic spacing. A qualitative chemical analysis based on the Al-$L$ edge showed an increase in the Al signal for the dark shear band segments and a decrease for the bright ones [Fig.~4(d), Supplemental Material] which is consistent with our previous report~\cite{Rosner2014}. It is worth noting that the chemical changes observed are minor. They are typically balanced between Al and Fe. It was further shown that the Al enrichment and the corresponding Fe depletion were not sufficient to account for the total density change in the dark shear band segments.The dominating effect causing the density change was the variation in free volume. Based on the fact that Fe is not soluble in fcc Al, we can estimate the maximum density variation arising from compositional changes. If we replace all of the present Fe (5~at\%) by Al, it would account for a maximum density decrease of roughly 4.5~\% which is not enough to explain a density decrease of 9~\%. Thus, the density drop for the dark shear band segments must include changes in free volume in addition to the changes in composition~\cite{Rosner2014}. Further information can be obtained from the zero lossand plasmon signal extracted from the EELS data. The increased zero losssignal [Fig.~4(b)] for the dark shear band regions indicates less scattering events due to less dense material at a constant foil thickness. On the other hand, the plasmon peak [Fig.~4(c)] reveals an increase for such dark shear band segments, which also indicates that compositional changes (Al enrichment) must have occurred, since an Al increase enhances the free-electron density and thus the plasmon signal.

The observation of an increased EELS ZLP for the dark shear band regions [see Supplemental Material Fig.~4(a)] and enhanced diffusion along shear bands by more than 6 to 8 orders of magnitude compared to diffusion in the undeformed matrix~\cite{Bokeloh2011} are perfectly consistent with enhanced free volume in the shear bands.

Structural changes inside shear bands have been analyzed indirectly by low-temperature heat capacity measurements on severely deformed metallic glasses~\cite{Bunz2014, Mitrofanov2014}. The so-called Boson peak revealed that the atomic structure was only modified inside the shear bands. Additional studies of the aging dependence of the Boson peak indicated that two different regions exist in the deformed specimens. One of these regions contributes to an accelerated aging of states since after identical relaxation treatment between room temperature and the glass transition, a state with lower enthalpy was attained by the deformed glass as compared to material without shear bands. Accelerated aging would be expected for the low-density regions of the shear band.

The HAADF image intensity shows no evidence for the formation of voids, but at this sample thickness it is insensitive to the small (1 nm diameter) voids previously reported~\cite{Li2002}. Reasons for void formation are the presence of multiple deflections of the shear bands, stress concentrators and low-viscosity regions. Nanovoids would also explain the early and catastrophic failure as well as the observation of vein patterns~\cite{Lewandowski2005,Greer2013} at the fracture surfaces of metallic glasses.

The repeated switching between higher and lower density observed along the shear band is directly analogous to the deformation behavior of granular media. Fazekas \textit{et al.}~\cite{Fazekas2007} have  shown by simulations that deformation of idealized granular media composed of spheres led to the formation of shear bands with high and low density regions. They further showed that the critical density of the shear bands depends on friction which, in turn, defines a specific packing state. Thus the theoretical findings for granular media explain our observation of higher and lower density shear band regions as the result of friction causing faster and slower motion of the shear bands. However, in the absence of granules, what plays their role in the metallic glass? It could be atoms, but the persistence of a similar structure of Al-like regions in a more disordered matrix within both the dark and bright segments leads us to suggest that ``the granules'' may instead be tightly bound short- or medium-range atomic clusters, which can fill the space more or less efficiently~\cite{Miracle2003, Miracle2004,Sheng2006}. Shear alters the packing of those clusters, but does not strongly modify their internal structure.

Since glasses do not have a defined slip plane it is expected that shear bands have a distribution of topological minima and maxima (``hills and valleys'') along their propagation direction, arising from inhomogeneity inherent to the glass. This nonplanar topology would give rise to a distribution of activation barriers for continued slip along the shear band. It is thus feasible, as a hypothesis, that compressive regions evolve before a local topological maximum and dilated regions may form after a topological maximum has been surmounted. Segments approaching a local barrier caused by a topological maximum would propagate slower while the motion speed would increase again after passing such a barrier resulting in a stick slip motion. This stick slip motion causes, in our opinion, the observed deflections. This interpretation fits very well with the observation in granular media~\cite{Fazekas2007} insomuch as the propagation velocity defines the packing state.

In summary, we have quantified the density inside a shear band of deformed Al$_{88}$Y$_7$Fe$_5$ metallic glass along its propagation direction. The local density varies enormously, ranging from -9~\% to +6~\%, compared to the undeformed matrix. Slight deflections from the main propagation direction of the shear band are found to coincide with the density switching from positive to negative values. The density alterations are attributed to compositional and free volume changes reflecting a different dense packing of atomic clusters, probably caused by different propagation speeds. These findings bear a striking resemblance to granular shear bands formed upon deformation of granular media~\cite{Fazekas2007}.

\begin{acknowledgments}
We kindly acknowledge financial support by the DFG via SPP 1594 (Topological engineering of ultra-strong glasses) and the U.S. National Science Foundation (DMR-1205899). We thank the Ernst Ruska-Centrum, Jülich, for the use of the Titan 80-300 STEM.\end{acknowledgments}

\bibliography{literature}

\begin{thebibliography}{39}%
\makeatletter
\providecommand \@ifxundefined [1]{%
 \@ifx{#1\undefined}
}%
\providecommand \@ifnum [1]{%
 \ifnum #1\expandafter \@firstoftwo
 \else \expandafter \@secondoftwo
 \fi
}%
\providecommand \@ifx [1]{%
 \ifx #1\expandafter \@firstoftwo
 \else \expandafter \@secondoftwo
 \fi
}%
\providecommand \natexlab [1]{#1}%
\providecommand \enquote  [1]{``#1''}%
\providecommand \bibnamefont  [1]{#1}%
\providecommand \bibfnamefont [1]{#1}%
\providecommand \citenamefont [1]{#1}%
\providecommand \href@noop [0]{\@secondoftwo}%
\providecommand \href [0]{\begingroup \@sanitize@url \@href}%
\providecommand \@href[1]{\@@startlink{#1}\@@href}%
\providecommand \@@href[1]{\endgroup#1\@@endlink}%
\providecommand \@sanitize@url [0]{\catcode `\\12\catcode `\$12\catcode
  `\&12\catcode `\#12\catcode `\^12\catcode `\_12\catcode `\%12\relax}%
\providecommand \@@startlink[1]{}%
\providecommand \@@endlink[0]{}%
\providecommand \url  [0]{\begingroup\@sanitize@url \@url }%
\providecommand \@url [1]{\endgroup\@href {#1}{\urlprefix }}%
\providecommand \urlprefix  [0]{URL }%
\providecommand \Eprint [0]{\href }%
\providecommand \doibase [0]{http://dx.doi.org/}%
\providecommand \selectlanguage [0]{\@gobble}%
\providecommand \bibinfo  [0]{\@secondoftwo}%
\providecommand \bibfield  [0]{\@secondoftwo}%
\providecommand \translation [1]{[#1]}%
\providecommand \BibitemOpen [0]{}%
\providecommand \bibitemStop [0]{}%
\providecommand \bibitemNoStop [0]{.\EOS\space}%
\providecommand \EOS [0]{\spacefactor3000\relax}%
\providecommand \BibitemShut  [1]{\csname bibitem#1\endcsname}%
\let\auto@bib@innerbib\@empty
\bibitem [{\citenamefont {Ashby}\ and\ \citenamefont
  {Greer}(2006)}]{Ashby2006}%
  \BibitemOpen
  \bibfield  {author} {\bibinfo {author} {\bibfnamefont {M.}~\bibnamefont
  {Ashby}}\ and\ \bibinfo {author} {\bibfnamefont {A.}~\bibnamefont {Greer}},\
  }\href {\doibase 10.1016/j.scriptamat.2005.09.051} {\bibfield  {journal}
  {\bibinfo  {journal} {Scripta Materialia}\ }\textbf {\bibinfo {volume}
  {54}},\ \bibinfo {pages} {321} (\bibinfo {year} {2006})}\BibitemShut
  {NoStop}%
\bibitem [{\citenamefont {Hays}\ \emph {et~al.}(2000)\citenamefont {Hays},
  \citenamefont {Kim},\ and\ \citenamefont {Johnson}}]{Hays2000}%
  \BibitemOpen
  \bibfield  {author} {\bibinfo {author} {\bibfnamefont {C.~C.}\ \bibnamefont
  {Hays}}, \bibinfo {author} {\bibfnamefont {C.~P.}\ \bibnamefont {Kim}}, \
  and\ \bibinfo {author} {\bibfnamefont {W.~L.}\ \bibnamefont {Johnson}},\
  }\href {\doibase 10.1103/PhysRevLett.84.2901} {\bibfield  {journal} {\bibinfo
   {journal} {Phys. Rev. Lett.}\ }\textbf {\bibinfo {volume} {84}},\ \bibinfo
  {pages} {2901} (\bibinfo {year} {2000})}\BibitemShut {NoStop}%
\bibitem [{\citenamefont {Szuecs}\ \emph {et~al.}(2001)\citenamefont {Szuecs},
  \citenamefont {Kim},\ and\ \citenamefont {Johnson}}]{Szuecs2001}%
  \BibitemOpen
  \bibfield  {author} {\bibinfo {author} {\bibfnamefont {F.}~\bibnamefont
  {Szuecs}}, \bibinfo {author} {\bibfnamefont {C.}~\bibnamefont {Kim}}, \ and\
  \bibinfo {author} {\bibfnamefont {W.}~\bibnamefont {Johnson}},\ }\href
  {\doibase http://dx.doi.org/10.1016/S1359-6454(01)00068-4} {\bibfield
  {journal} {\bibinfo  {journal} {Acta Materialia}\ }\textbf {\bibinfo {volume}
  {49}},\ \bibinfo {pages} {1507 } (\bibinfo {year} {2001})}\BibitemShut
  {NoStop}%
\bibitem [{\citenamefont {Donohue}\ \emph {et~al.}(2007)\citenamefont
  {Donohue}, \citenamefont {Spaepen}, \citenamefont {Hoagland},\ and\
  \citenamefont {Misra}}]{Donohue2007}%
  \BibitemOpen
  \bibfield  {author} {\bibinfo {author} {\bibfnamefont {A.}~\bibnamefont
  {Donohue}}, \bibinfo {author} {\bibfnamefont {F.}~\bibnamefont {Spaepen}},
  \bibinfo {author} {\bibfnamefont {R.~G.}\ \bibnamefont {Hoagland}}, \ and\
  \bibinfo {author} {\bibfnamefont {A.}~\bibnamefont {Misra}},\ }\href
  {\doibase 10.1063/1.2821227} {\bibfield  {journal} {\bibinfo  {journal}
  {Applied Physics Letters}\ }\textbf {\bibinfo {volume} {91}},\ \bibinfo
  {pages} {241905} (\bibinfo {year} {2007})}\BibitemShut {NoStop}%
\bibitem [{\citenamefont {Eckert}\ \emph {et~al.}(2007)\citenamefont {Eckert},
  \citenamefont {Das}, \citenamefont {Pauly},\ and\ \citenamefont
  {Duhamel}}]{Eckert2007}%
  \BibitemOpen
  \bibfield  {author} {\bibinfo {author} {\bibfnamefont {J.}~\bibnamefont
  {Eckert}}, \bibinfo {author} {\bibfnamefont {J.}~\bibnamefont {Das}},
  \bibinfo {author} {\bibfnamefont {S.}~\bibnamefont {Pauly}}, \ and\ \bibinfo
  {author} {\bibfnamefont {C.}~\bibnamefont {Duhamel}},\ }\href {\doibase
  10.1557/jmr.2007.0050} {\bibfield  {journal} {\bibinfo  {journal} {Journal of
  Materials Research}\ }\textbf {\bibinfo {volume} {22}},\ \bibinfo {pages}
  {285} (\bibinfo {year} {2007})}\BibitemShut {NoStop}%
\bibitem [{\citenamefont {Hofmann}\ \emph {et~al.}(2008)\citenamefont
  {Hofmann}, \citenamefont {Suh}, \citenamefont {Wiest}, \citenamefont {Duan},
  \citenamefont {Lind}, \citenamefont {Demetriou},\ and\ \citenamefont
  {Johnson}}]{Hofmann2008}%
  \BibitemOpen
  \bibfield  {author} {\bibinfo {author} {\bibfnamefont {D.~C.}\ \bibnamefont
  {Hofmann}}, \bibinfo {author} {\bibfnamefont {J.-Y.}\ \bibnamefont {Suh}},
  \bibinfo {author} {\bibfnamefont {A.}~\bibnamefont {Wiest}}, \bibinfo
  {author} {\bibfnamefont {G.}~\bibnamefont {Duan}}, \bibinfo {author}
  {\bibfnamefont {M.-L.}\ \bibnamefont {Lind}}, \bibinfo {author}
  {\bibfnamefont {M.~D.}\ \bibnamefont {Demetriou}}, \ and\ \bibinfo {author}
  {\bibfnamefont {W.~L.}\ \bibnamefont {Johnson}},\ }\href {\doibase
  10.1038/nature06598} {\bibfield  {journal} {\bibinfo  {journal} {Nature}\
  }\textbf {\bibinfo {volume} {451}},\ \bibinfo {pages} {1085} (\bibinfo {year}
  {2008})}\BibitemShut {NoStop}%
\bibitem [{\citenamefont {Schroers}\ and\ \citenamefont
  {Johnson}(2004)}]{Schroers2004}%
  \BibitemOpen
  \bibfield  {author} {\bibinfo {author} {\bibfnamefont {J.}~\bibnamefont
  {Schroers}}\ and\ \bibinfo {author} {\bibfnamefont {W.~L.}\ \bibnamefont
  {Johnson}},\ }\href {\doibase 10.1103/PhysRevLett.93.255506} {\bibfield
  {journal} {\bibinfo  {journal} {Phys. Rev. Lett.}\ }\textbf {\bibinfo
  {volume} {93}},\ \bibinfo {pages} {255506} (\bibinfo {year}
  {2004})}\BibitemShut {NoStop}%
\bibitem [{\citenamefont {Lewandowski}\ \emph {et~al.}(2005)\citenamefont
  {Lewandowski}, \citenamefont {Wang},\ and\ \citenamefont
  {Greer}}]{Lewandowski2005}%
  \BibitemOpen
  \bibfield  {author} {\bibinfo {author} {\bibfnamefont {J.~J.}\ \bibnamefont
  {Lewandowski}}, \bibinfo {author} {\bibfnamefont {W.~H.}\ \bibnamefont
  {Wang}}, \ and\ \bibinfo {author} {\bibfnamefont {A.~L.}\ \bibnamefont
  {Greer}},\ }\href {\doibase 10.1080/09500830500080474} {\bibfield  {journal}
  {\bibinfo  {journal} {Philosophical Magazine Letters}\ }\textbf {\bibinfo
  {volume} {85}},\ \bibinfo {pages} {77} (\bibinfo {year} {2005})}\BibitemShut
  {NoStop}%
\bibitem [{\citenamefont {Demetriou}\ \emph {et~al.}(2011)\citenamefont
  {Demetriou}, \citenamefont {Launey}, \citenamefont {Garrett}, \citenamefont
  {Schramm}, \citenamefont {Hofmann}, \citenamefont {Johnson},\ and\
  \citenamefont {Ritchie}}]{Demetriou2011}%
  \BibitemOpen
  \bibfield  {author} {\bibinfo {author} {\bibfnamefont {M.~D.}\ \bibnamefont
  {Demetriou}}, \bibinfo {author} {\bibfnamefont {M.~E.}\ \bibnamefont
  {Launey}}, \bibinfo {author} {\bibfnamefont {G.}~\bibnamefont {Garrett}},
  \bibinfo {author} {\bibfnamefont {J.~P.}\ \bibnamefont {Schramm}}, \bibinfo
  {author} {\bibfnamefont {D.~C.}\ \bibnamefont {Hofmann}}, \bibinfo {author}
  {\bibfnamefont {W.~L.}\ \bibnamefont {Johnson}}, \ and\ \bibinfo {author}
  {\bibfnamefont {R.~O.}\ \bibnamefont {Ritchie}},\ }\href {\doibase
  10.1038/nmat2930} {\bibfield  {journal} {\bibinfo  {journal} {Nature
  materials}\ }\textbf {\bibinfo {volume} {10}},\ \bibinfo {pages} {123}
  (\bibinfo {year} {2011})}\BibitemShut {NoStop}%
\bibitem [{\citenamefont {Wang}\ \emph {et~al.}(2011)\citenamefont {Wang},
  \citenamefont {Cao}, \citenamefont {Chen}, \citenamefont {Hono},
  \citenamefont {Zhong}, \citenamefont {Jiang}, \citenamefont {Nie},
  \citenamefont {Chen}, \citenamefont {Wang},\ and\ \citenamefont
  {Jiang}}]{Wang2011}%
  \BibitemOpen
  \bibfield  {author} {\bibinfo {author} {\bibfnamefont {X.}~\bibnamefont
  {Wang}}, \bibinfo {author} {\bibfnamefont {Q.~P.}\ \bibnamefont {Cao}},
  \bibinfo {author} {\bibfnamefont {Y.~M.}\ \bibnamefont {Chen}}, \bibinfo
  {author} {\bibfnamefont {K.}~\bibnamefont {Hono}}, \bibinfo {author}
  {\bibfnamefont {C.}~\bibnamefont {Zhong}}, \bibinfo {author} {\bibfnamefont
  {Q.~K.}\ \bibnamefont {Jiang}}, \bibinfo {author} {\bibfnamefont {X.~P.}\
  \bibnamefont {Nie}}, \bibinfo {author} {\bibfnamefont {L.~Y.}\ \bibnamefont
  {Chen}}, \bibinfo {author} {\bibfnamefont {X.~D.}\ \bibnamefont {Wang}}, \
  and\ \bibinfo {author} {\bibfnamefont {J.~Z.}\ \bibnamefont {Jiang}},\ }\href
  {\doibase 10.1016/j.actamat.2010.10.034} {\bibfield  {journal} {\bibinfo
  {journal} {Acta Materialia}\ }\textbf {\bibinfo {volume} {59}},\ \bibinfo
  {pages} {1037} (\bibinfo {year} {2011})}\BibitemShut {NoStop}%
\bibitem [{\citenamefont {Spaepen}(1977)}]{Spaepen1977}%
  \BibitemOpen
  \bibfield  {author} {\bibinfo {author} {\bibfnamefont {F.}~\bibnamefont
  {Spaepen}},\ }\href {\doibase 10.1016/0001-6160(77)90232-2} {\bibfield
  {journal} {\bibinfo  {journal} {Acta Metallurgica}\ }\textbf {\bibinfo
  {volume} {25}},\ \bibinfo {pages} {407} (\bibinfo {year} {1977})}\BibitemShut
  {NoStop}%
\bibitem [{\citenamefont {Argon}(1979)}]{Argon1979}%
  \BibitemOpen
  \bibfield  {author} {\bibinfo {author} {\bibfnamefont {A.}~\bibnamefont
  {Argon}},\ }\href {\doibase 10.1016/0001-6160(79)90055-5} {\bibfield
  {journal} {\bibinfo  {journal} {Acta Metallurgica}\ }\textbf {\bibinfo
  {volume} {27}},\ \bibinfo {pages} {47} (\bibinfo {year} {1979})}\BibitemShut
  {NoStop}%
\bibitem [{\citenamefont {Greer}\ \emph {et~al.}(2013)\citenamefont {Greer},
  \citenamefont {Cheng},\ and\ \citenamefont {Ma}}]{Greer2013}%
  \BibitemOpen
  \bibfield  {author} {\bibinfo {author} {\bibfnamefont {A.}~\bibnamefont
  {Greer}}, \bibinfo {author} {\bibfnamefont {Y.}~\bibnamefont {Cheng}}, \ and\
  \bibinfo {author} {\bibfnamefont {E.}~\bibnamefont {Ma}},\ }\href {\doibase
  10.1016/j.mser.2013.04.001} {\bibfield  {journal} {\bibinfo  {journal}
  {Materials Science and Engineering: R: Reports}\ }\textbf {\bibinfo {volume}
  {74}},\ \bibinfo {pages} {71} (\bibinfo {year} {2013})}\BibitemShut {NoStop}%
\bibitem [{\citenamefont {Donovan}\ and\ \citenamefont
  {Stobbs}(1981)}]{Donovan1981}%
  \BibitemOpen
  \bibfield  {author} {\bibinfo {author} {\bibfnamefont {P.}~\bibnamefont
  {Donovan}}\ and\ \bibinfo {author} {\bibfnamefont {W.}~\bibnamefont
  {Stobbs}},\ }\href {\doibase 10.1016/0001-6160(81)90177-2} {\bibfield
  {journal} {\bibinfo  {journal} {Acta Metallurgica}\ }\textbf {\bibinfo
  {volume} {29}},\ \bibinfo {pages} {1419} (\bibinfo {year}
  {1981})}\BibitemShut {NoStop}%
\bibitem [{\citenamefont {Klaum\"unzer}\ \emph {et~al.}(2011)\citenamefont
  {Klaum\"unzer}, \citenamefont {Lazarev}, \citenamefont {Maa\ss{}},
  \citenamefont {Dalla~Torre}, \citenamefont {Vinogradov},\ and\ \citenamefont
  {L\"offler}}]{Klaumunzer2011}%
  \BibitemOpen
  \bibfield  {author} {\bibinfo {author} {\bibfnamefont {D.}~\bibnamefont
  {Klaum\"unzer}}, \bibinfo {author} {\bibfnamefont {A.}~\bibnamefont
  {Lazarev}}, \bibinfo {author} {\bibfnamefont {R.}~\bibnamefont {Maa\ss{}}},
  \bibinfo {author} {\bibfnamefont {F.~H.}\ \bibnamefont {Dalla~Torre}},
  \bibinfo {author} {\bibfnamefont {A.}~\bibnamefont {Vinogradov}}, \ and\
  \bibinfo {author} {\bibfnamefont {J.~F.}\ \bibnamefont {L\"offler}},\ }\href
  {\doibase 10.1103/PhysRevLett.107.185502} {\bibfield  {journal} {\bibinfo
  {journal} {Phys. Rev. Lett.}\ }\textbf {\bibinfo {volume} {107}},\ \bibinfo
  {pages} {185502} (\bibinfo {year} {2011})}\BibitemShut {NoStop}%
\bibitem [{\citenamefont {Pan}\ \emph {et~al.}(2011)\citenamefont {Pan},
  \citenamefont {Chen}, \citenamefont {Liu},\ and\ \citenamefont
  {Li}}]{Pan2011}%
  \BibitemOpen
  \bibfield  {author} {\bibinfo {author} {\bibfnamefont {J.}~\bibnamefont
  {Pan}}, \bibinfo {author} {\bibfnamefont {Q.}~\bibnamefont {Chen}}, \bibinfo
  {author} {\bibfnamefont {L.}~\bibnamefont {Liu}}, \ and\ \bibinfo {author}
  {\bibfnamefont {Y.}~\bibnamefont {Li}},\ }\href {\doibase
  10.1016/j.actamat.2011.04.047} {\bibfield  {journal} {\bibinfo  {journal}
  {Acta Materialia}\ }\textbf {\bibinfo {volume} {59}},\ \bibinfo {pages}
  {5146} (\bibinfo {year} {2011})}\BibitemShut {NoStop}%
\bibitem [{\citenamefont {Nagel}\ \emph {et~al.}(1998)\citenamefont {Nagel},
  \citenamefont {R\"{a}tzke}, \citenamefont {Schmidtke}, \citenamefont {Wolff},
  \citenamefont {Geyer},\ and\ \citenamefont {Faupel}}]{Nagel1998}%
  \BibitemOpen
  \bibfield  {author} {\bibinfo {author} {\bibfnamefont {C.}~\bibnamefont
  {Nagel}}, \bibinfo {author} {\bibfnamefont {K.}~\bibnamefont {R\"{a}tzke}},
  \bibinfo {author} {\bibfnamefont {E.}~\bibnamefont {Schmidtke}}, \bibinfo
  {author} {\bibfnamefont {J.}~\bibnamefont {Wolff}}, \bibinfo {author}
  {\bibfnamefont {U.}~\bibnamefont {Geyer}}, \ and\ \bibinfo {author}
  {\bibfnamefont {F.}~\bibnamefont {Faupel}},\ }\href {\doibase
  10.1103/PhysRevB.57.10224} {\bibfield  {journal} {\bibinfo  {journal}
  {Physical Review B}\ }\textbf {\bibinfo {volume} {57}},\ \bibinfo {pages}
  {10224} (\bibinfo {year} {1998})}\BibitemShut {NoStop}%
\bibitem [{\citenamefont {Wilde}\ \emph {et~al.}(2000)\citenamefont {Wilde},
  \citenamefont {G\"orler}, \citenamefont {Willnecker},\ and\ \citenamefont
  {Fecht}}]{Wilde2000}%
  \BibitemOpen
  \bibfield  {author} {\bibinfo {author} {\bibfnamefont {G.}~\bibnamefont
  {Wilde}}, \bibinfo {author} {\bibfnamefont {G.~P.}\ \bibnamefont {G\"orler}},
  \bibinfo {author} {\bibfnamefont {R.}~\bibnamefont {Willnecker}}, \ and\
  \bibinfo {author} {\bibfnamefont {H.~J.}\ \bibnamefont {Fecht}},\ }\href
  {\doibase 10.1063/1.371991} {\bibfield  {journal} {\bibinfo  {journal}
  {Journal of Applied Physics}\ }\textbf {\bibinfo {volume} {87}},\ \bibinfo
  {pages} {1141} (\bibinfo {year} {2000})}\BibitemShut {NoStop}%
\bibitem [{\citenamefont {Chen}\ \emph {et~al.}(2009)\citenamefont {Chen},
  \citenamefont {Ohkubo}, \citenamefont {Mukai},\ and\ \citenamefont
  {Hono}}]{Chen2009}%
  \BibitemOpen
  \bibfield  {author} {\bibinfo {author} {\bibfnamefont {Y.}~\bibnamefont
  {Chen}}, \bibinfo {author} {\bibfnamefont {T.}~\bibnamefont {Ohkubo}},
  \bibinfo {author} {\bibfnamefont {T.}~\bibnamefont {Mukai}}, \ and\ \bibinfo
  {author} {\bibfnamefont {K.}~\bibnamefont {Hono}},\ }\href {\doibase
  10.1557/JMR.2009.0001} {\bibfield  {journal} {\bibinfo  {journal} {Journal of
  Materials Research}\ }\textbf {\bibinfo {volume} {24}},\ \bibinfo {pages} {1}
  (\bibinfo {year} {2009})}\BibitemShut {NoStop}%
\bibitem [{\citenamefont {Zhang}\ and\ \citenamefont {Hahn}(2009)}]{Zhang2009}%
  \BibitemOpen
  \bibfield  {author} {\bibinfo {author} {\bibfnamefont {Y.}~\bibnamefont
  {Zhang}}\ and\ \bibinfo {author} {\bibfnamefont {H.}~\bibnamefont {Hahn}},\
  }\href {\doibase 10.1016/j.jallcom.2009.08.091} {\bibfield  {journal}
  {\bibinfo  {journal} {Journal of Alloys and Compounds}\ }\textbf {\bibinfo
  {volume} {488}},\ \bibinfo {pages} {65} (\bibinfo {year} {2009})}\BibitemShut
  {NoStop}%
\bibitem [{\citenamefont {Dmowski}\ \emph {et~al.}(2010)\citenamefont
  {Dmowski}, \citenamefont {Iwashita}, \citenamefont {Chuang}, \citenamefont
  {Almer},\ and\ \citenamefont {Egami}}]{Dmowski2010}%
  \BibitemOpen
  \bibfield  {author} {\bibinfo {author} {\bibfnamefont {W.}~\bibnamefont
  {Dmowski}}, \bibinfo {author} {\bibfnamefont {T.}~\bibnamefont {Iwashita}},
  \bibinfo {author} {\bibfnamefont {C.-P.}\ \bibnamefont {Chuang}}, \bibinfo
  {author} {\bibfnamefont {J.}~\bibnamefont {Almer}}, \ and\ \bibinfo {author}
  {\bibfnamefont {T.}~\bibnamefont {Egami}},\ }\href {\doibase
  10.1103/PhysRevLett.105.205502} {\bibfield  {journal} {\bibinfo  {journal}
  {Physical Review Letters}\ }\textbf {\bibinfo {volume} {105}},\ \bibinfo
  {pages} {205502} (\bibinfo {year} {2010})}\BibitemShut {NoStop}%
\bibitem [{\citenamefont {Lechner}\ \emph {et~al.}(2010)\citenamefont
  {Lechner}, \citenamefont {Puff}, \citenamefont {Wilde},\ and\ \citenamefont
  {W\"{u}rschum}}]{Lechner2010}%
  \BibitemOpen
  \bibfield  {author} {\bibinfo {author} {\bibfnamefont {W.}~\bibnamefont
  {Lechner}}, \bibinfo {author} {\bibfnamefont {W.}~\bibnamefont {Puff}},
  \bibinfo {author} {\bibfnamefont {G.}~\bibnamefont {Wilde}}, \ and\ \bibinfo
  {author} {\bibfnamefont {R.}~\bibnamefont {W\"{u}rschum}},\ }\href {\doibase
  10.1016/j.scriptamat.2009.11.037} {\bibfield  {journal} {\bibinfo  {journal}
  {Scripta Materialia}\ }\textbf {\bibinfo {volume} {62}},\ \bibinfo {pages}
  {439} (\bibinfo {year} {2010})}\BibitemShut {NoStop}%
\bibitem [{\citenamefont {Shao}\ \emph {et~al.}(2013)\citenamefont {Shao},
  \citenamefont {Xu}, \citenamefont {Shi}, \citenamefont {Yu}, \citenamefont
  {Hahn}, \citenamefont {Gleiter},\ and\ \citenamefont {Li}}]{Shao2013}%
  \BibitemOpen
  \bibfield  {author} {\bibinfo {author} {\bibfnamefont {H.}~\bibnamefont
  {Shao}}, \bibinfo {author} {\bibfnamefont {Y.}~\bibnamefont {Xu}}, \bibinfo
  {author} {\bibfnamefont {B.}~\bibnamefont {Shi}}, \bibinfo {author}
  {\bibfnamefont {C.}~\bibnamefont {Yu}}, \bibinfo {author} {\bibfnamefont
  {H.}~\bibnamefont {Hahn}}, \bibinfo {author} {\bibfnamefont {H.}~\bibnamefont
  {Gleiter}}, \ and\ \bibinfo {author} {\bibfnamefont {J.}~\bibnamefont {Li}},\
  }\href {\doibase 10.1016/j.jallcom.2012.08.132} {\bibfield  {journal}
  {\bibinfo  {journal} {Journal of Alloys and Compounds}\ }\textbf {\bibinfo
  {volume} {548}},\ \bibinfo {pages} {77} (\bibinfo {year} {2013})}\BibitemShut
  {NoStop}%
\bibitem [{\citenamefont {Li}\ \emph {et~al.}(2002)\citenamefont {Li},
  \citenamefont {Spaepen},\ and\ \citenamefont {Hufnagel}}]{Li2002}%
  \BibitemOpen
  \bibfield  {author} {\bibinfo {author} {\bibfnamefont {J.}~\bibnamefont
  {Li}}, \bibinfo {author} {\bibfnamefont {F.}~\bibnamefont {Spaepen}}, \ and\
  \bibinfo {author} {\bibfnamefont {T.~C.}\ \bibnamefont {Hufnagel}},\ }\href
  {\doibase 10.1080/01418610210152792} {\bibfield  {journal} {\bibinfo
  {journal} {Philosophical Magazine A}\ }\textbf {\bibinfo {volume} {82}},\
  \bibinfo {pages} {2623} (\bibinfo {year} {2002})}\BibitemShut {NoStop}%
\bibitem [{\citenamefont {R\"{o}sner}\ \emph {et~al.}(2014)\citenamefont
  {R\"{o}sner}, \citenamefont {Peterlechner}, \citenamefont {K\"{u}bel},
  \citenamefont {Schmidt},\ and\ \citenamefont {Wilde}}]{Rosner2014}%
  \BibitemOpen
  \bibfield  {author} {\bibinfo {author} {\bibfnamefont {H.}~\bibnamefont
  {R\"{o}sner}}, \bibinfo {author} {\bibfnamefont {M.}~\bibnamefont
  {Peterlechner}}, \bibinfo {author} {\bibfnamefont {C.}~\bibnamefont
  {K\"{u}bel}}, \bibinfo {author} {\bibfnamefont {V.}~\bibnamefont {Schmidt}},
  \ and\ \bibinfo {author} {\bibfnamefont {G.}~\bibnamefont {Wilde}},\ }\href
  {\doibase 10.1016/j.ultramic.2014.03.006} {\bibfield  {journal} {\bibinfo
  {journal} {Ultramicroscopy}\ }\textbf {\bibinfo {volume} {142}},\ \bibinfo
  {pages} {1} (\bibinfo {year} {2014})}\BibitemShut {NoStop}%
\bibitem [{\citenamefont {Wilde}\ \emph {et~al.}(2003)\citenamefont {Wilde},
  \citenamefont {Boucharat}, \citenamefont {Hebert}, \citenamefont
  {R\"{o}sner}, \citenamefont {Tong},\ and\ \citenamefont
  {Perepezko}}]{Wilde2003}%
  \BibitemOpen
  \bibfield  {author} {\bibinfo {author} {\bibfnamefont {G.}~\bibnamefont
  {Wilde}}, \bibinfo {author} {\bibfnamefont {N.}~\bibnamefont {Boucharat}},
  \bibinfo {author} {\bibfnamefont {R.}~\bibnamefont {Hebert}}, \bibinfo
  {author} {\bibfnamefont {H.}~\bibnamefont {R\"{o}sner}}, \bibinfo {author}
  {\bibfnamefont {W.}~\bibnamefont {Tong}}, \ and\ \bibinfo {author}
  {\bibfnamefont {J.}~\bibnamefont {Perepezko}},\ }\href {\doibase
  10.1002/adem.200390019} {\bibfield  {journal} {\bibinfo  {journal} {Advanced
  Engineering Materials}\ }\textbf {\bibinfo {volume} {5}},\ \bibinfo {pages}
  {125} (\bibinfo {year} {2003})}\BibitemShut {NoStop}%
\bibitem [{\citenamefont {Daulton}\ \emph {et~al.}(2010)\citenamefont
  {Daulton}, \citenamefont {Bondi},\ and\ \citenamefont
  {Kelton}}]{Daulton2010}%
  \BibitemOpen
  \bibfield  {author} {\bibinfo {author} {\bibfnamefont {T.}~\bibnamefont
  {Daulton}}, \bibinfo {author} {\bibfnamefont {K.}~\bibnamefont {Bondi}}, \
  and\ \bibinfo {author} {\bibfnamefont {K.}~\bibnamefont {Kelton}},\ }\href
  {\doibase 10.1016/j.ultramic.2010.05.010} {\bibfield  {journal} {\bibinfo
  {journal} {Ultramicroscopy}\ }\textbf {\bibinfo {volume} {110}},\ \bibinfo
  {pages} {1279} (\bibinfo {year} {2010})}\BibitemShut {NoStop}%
\bibitem [{\citenamefont {Yi}\ and\ \citenamefont {Voyles}(2012)}]{Yi2012}%
  \BibitemOpen
  \bibfield  {author} {\bibinfo {author} {\bibfnamefont {F.}~\bibnamefont
  {Yi}}\ and\ \bibinfo {author} {\bibfnamefont {P.~M.}\ \bibnamefont
  {Voyles}},\ }\href {\doibase 10.1016/j.ultramic.2012.07.022} {\bibfield
  {journal} {\bibinfo  {journal} {Ultramicroscopy}\ }\textbf {\bibinfo {volume}
  {122}},\ \bibinfo {pages} {37} (\bibinfo {year} {2012})}\BibitemShut
  {NoStop}%
\bibitem [{\citenamefont {Boucharat}\ \emph {et~al.}(2005)\citenamefont
  {Boucharat}, \citenamefont {Hebert}, \citenamefont {R\"{o}sner},
  \citenamefont {Valiev},\ and\ \citenamefont {Wilde}}]{Boucharat2005}%
  \BibitemOpen
  \bibfield  {author} {\bibinfo {author} {\bibfnamefont {N.}~\bibnamefont
  {Boucharat}}, \bibinfo {author} {\bibfnamefont {R.}~\bibnamefont {Hebert}},
  \bibinfo {author} {\bibfnamefont {H.}~\bibnamefont {R\"{o}sner}}, \bibinfo
  {author} {\bibfnamefont {R.}~\bibnamefont {Valiev}}, \ and\ \bibinfo {author}
  {\bibfnamefont {G.}~\bibnamefont {Wilde}},\ }\href {\doibase
  10.1016/j.scriptamat.2005.06.004} {\bibfield  {journal} {\bibinfo  {journal}
  {Scripta Materialia}\ }\textbf {\bibinfo {volume} {53}},\ \bibinfo {pages}
  {823} (\bibinfo {year} {2005})}\BibitemShut {NoStop}%
\bibitem [{\citenamefont {Hirata}\ and\ \citenamefont
  {Chen}(2014)}]{Hirata2014}%
  \BibitemOpen
  \bibfield  {author} {\bibinfo {author} {\bibfnamefont {A.}~\bibnamefont
  {Hirata}}\ and\ \bibinfo {author} {\bibfnamefont {M.}~\bibnamefont {Chen}},\
  }\href {\doibase 10.1016/j.jnoncrysol.2013.03.010} {\bibfield  {journal}
  {\bibinfo  {journal} {Journal of Non-Crystalline Solids}\ }\textbf {\bibinfo
  {volume} {383}},\ \bibinfo {pages} {52} (\bibinfo {year} {2014})}\BibitemShut
  {NoStop}%
\bibitem [{\citenamefont {Bokeloh}\ \emph {et~al.}(2011)\citenamefont
  {Bokeloh}, \citenamefont {Divinski}, \citenamefont {Reglitz},\ and\
  \citenamefont {Wilde}}]{Bokeloh2011}%
  \BibitemOpen
  \bibfield  {author} {\bibinfo {author} {\bibfnamefont {J.}~\bibnamefont
  {Bokeloh}}, \bibinfo {author} {\bibfnamefont {S.~V.}\ \bibnamefont
  {Divinski}}, \bibinfo {author} {\bibfnamefont {G.}~\bibnamefont {Reglitz}}, \
  and\ \bibinfo {author} {\bibfnamefont {G.}~\bibnamefont {Wilde}},\ }\href
  {\doibase 10.1103/PhysRevLett.107.235503} {\bibfield  {journal} {\bibinfo
  {journal} {Physical Review Letters}\ }\textbf {\bibinfo {volume} {107}},\
  \bibinfo {pages} {235503} (\bibinfo {year} {2011})}\BibitemShut {NoStop}%
\bibitem [{\citenamefont {B\"{u}nz}\ \emph {et~al.}(2014)\citenamefont
  {B\"{u}nz}, \citenamefont {Brink}, \citenamefont {Tsuchiya}, \citenamefont
  {Meng}, \citenamefont {Wilde},\ and\ \citenamefont {Albe}}]{Bunz2014}%
  \BibitemOpen
  \bibfield  {author} {\bibinfo {author} {\bibfnamefont {J.}~\bibnamefont
  {B\"{u}nz}}, \bibinfo {author} {\bibfnamefont {T.}~\bibnamefont {Brink}},
  \bibinfo {author} {\bibfnamefont {K.}~\bibnamefont {Tsuchiya}}, \bibinfo
  {author} {\bibfnamefont {F.}~\bibnamefont {Meng}}, \bibinfo {author}
  {\bibfnamefont {G.}~\bibnamefont {Wilde}}, \ and\ \bibinfo {author}
  {\bibfnamefont {K.}~\bibnamefont {Albe}},\ }\href {\doibase
  10.1103/PhysRevLett.112.135501} {\bibfield  {journal} {\bibinfo  {journal}
  {Physical Review Letters}\ }\textbf {\bibinfo {volume} {112}},\ \bibinfo
  {pages} {135501} (\bibinfo {year} {2014})}\BibitemShut {NoStop}%
\bibitem [{\citenamefont {Mitrofanov}\ \emph {et~al.}(2014)\citenamefont
  {Mitrofanov}, \citenamefont {Peterlechner}, \citenamefont {Divinski},\ and\
  \citenamefont {Wilde}}]{Mitrofanov2014}%
  \BibitemOpen
  \bibfield  {author} {\bibinfo {author} {\bibfnamefont {Y.~P.}\ \bibnamefont
  {Mitrofanov}}, \bibinfo {author} {\bibfnamefont {M.}~\bibnamefont
  {Peterlechner}}, \bibinfo {author} {\bibfnamefont {S.~V.}\ \bibnamefont
  {Divinski}}, \ and\ \bibinfo {author} {\bibfnamefont {G.}~\bibnamefont
  {Wilde}},\ }\href {\doibase 10.1103/PhysRevLett.112.135901} {\bibfield
  {journal} {\bibinfo  {journal} {Physical Review Letters}\ }\textbf {\bibinfo
  {volume} {112}},\ \bibinfo {pages} {135901} (\bibinfo {year}
  {2014})}\BibitemShut {NoStop}%
\bibitem [{\citenamefont {Fazekas}\ \emph {et~al.}(2007)\citenamefont
  {Fazekas}, \citenamefont {T\"{o}r\"{o}k},\ and\ \citenamefont
  {Kert\'{e}sz}}]{Fazekas2007}%
  \BibitemOpen
  \bibfield  {author} {\bibinfo {author} {\bibfnamefont {S.}~\bibnamefont
  {Fazekas}}, \bibinfo {author} {\bibfnamefont {J.}~\bibnamefont
  {T\"{o}r\"{o}k}}, \ and\ \bibinfo {author} {\bibfnamefont {J.}~\bibnamefont
  {Kert\'{e}sz}},\ }\href {\doibase 10.1103/PhysRevE.75.011302} {\bibfield
  {journal} {\bibinfo  {journal} {Physical Review E}\ }\textbf {\bibinfo
  {volume} {75}},\ \bibinfo {pages} {011302} (\bibinfo {year}
  {2007})}\BibitemShut {NoStop}%
\bibitem [{\citenamefont {Miracle}\ \emph {et~al.}(2003)\citenamefont
  {Miracle}, \citenamefont {Sanders},\ and\ \citenamefont
  {Senkov}}]{Miracle2003}%
  \BibitemOpen
  \bibfield  {author} {\bibinfo {author} {\bibfnamefont {D.}~\bibnamefont
  {Miracle}}, \bibinfo {author} {\bibfnamefont {W.}~\bibnamefont {Sanders}}, \
  and\ \bibinfo {author} {\bibfnamefont {O.~N.}\ \bibnamefont {Senkov}},\
  }\href {\doibase 10.1080/1478643031000098828} {\bibfield  {journal} {\bibinfo
   {journal} {Philosophical Magazine}\ }\textbf {\bibinfo {volume} {83}},\
  \bibinfo {pages} {2409} (\bibinfo {year} {2003})}\BibitemShut {NoStop}%
\bibitem [{\citenamefont {Miracle}(2004)}]{Miracle2004}%
  \BibitemOpen
  \bibfield  {author} {\bibinfo {author} {\bibfnamefont {D.~B.}\ \bibnamefont
  {Miracle}},\ }\href {\doibase 10.1038/nmat1219} {\bibfield  {journal}
  {\bibinfo  {journal} {Nature materials}\ }\textbf {\bibinfo {volume} {3}},\
  \bibinfo {pages} {697} (\bibinfo {year} {2004})}\BibitemShut {NoStop}%
\bibitem [{\citenamefont {Sheng}\ \emph {et~al.}(2006)\citenamefont {Sheng},
  \citenamefont {Luo}, \citenamefont {Alamgir}, \citenamefont {Bai},\ and\
  \citenamefont {Ma}}]{Sheng2006}%
  \BibitemOpen
  \bibfield  {author} {\bibinfo {author} {\bibfnamefont {H.~W.}\ \bibnamefont
  {Sheng}}, \bibinfo {author} {\bibfnamefont {W.~K.}\ \bibnamefont {Luo}},
  \bibinfo {author} {\bibfnamefont {F.~M.}\ \bibnamefont {Alamgir}}, \bibinfo
  {author} {\bibfnamefont {J.~M.}\ \bibnamefont {Bai}}, \ and\ \bibinfo
  {author} {\bibfnamefont {E.}~\bibnamefont {Ma}},\ }\href {\doibase
  10.1038/nature04421} {\bibfield  {journal} {\bibinfo  {journal} {Nature}\
  }\textbf {\bibinfo {volume} {439}},\ \bibinfo {pages} {419} (\bibinfo {year}
  {2006})}\BibitemShut {NoStop}%
\bibitem [{\citenamefont {Iakoubovskii}\ \emph {et~al.}(2008)\citenamefont
  {Iakoubovskii}, \citenamefont {Mitsuishi}, \citenamefont {Nakayama},\ and\
  \citenamefont {Furuya}}]{Iakoubovskii2008}%
  \BibitemOpen
  \bibfield  {author} {\bibinfo {author} {\bibfnamefont {K.}~\bibnamefont
  {Iakoubovskii}}, \bibinfo {author} {\bibfnamefont {K.}~\bibnamefont
  {Mitsuishi}}, \bibinfo {author} {\bibfnamefont {Y.}~\bibnamefont {Nakayama}},
  \ and\ \bibinfo {author} {\bibfnamefont {K.}~\bibnamefont {Furuya}},\ }\href
  {\doibase 10.1002/jemt.20597} {\bibfield  {journal} {\bibinfo  {journal}
  {Microscopy research and technique}\ }\textbf {\bibinfo {volume} {71}},\
  \bibinfo {pages} {626} (\bibinfo {year} {2008})}\BibitemShut {NoStop}%
\bibitem [{\citenamefont {Gammer}\ \emph {et~al.}(2010)\citenamefont {Gammer},
  \citenamefont {Mangler}, \citenamefont {Rentenberger},\ and\ \citenamefont
  {Karnthaler}}]{Gammer2010}%
  \BibitemOpen
  \bibfield  {author} {\bibinfo {author} {\bibfnamefont {C.}~\bibnamefont
  {Gammer}}, \bibinfo {author} {\bibfnamefont {C.}~\bibnamefont {Mangler}},
  \bibinfo {author} {\bibfnamefont {C.}~\bibnamefont {Rentenberger}}, \ and\
  \bibinfo {author} {\bibfnamefont {H.}~\bibnamefont {Karnthaler}},\ }\href
  {\doibase 10.1016/j.scriptamat.2010.04.019} {\bibfield  {journal} {\bibinfo
  {journal} {Scripta Materialia}\ }\textbf {\bibinfo {volume} {63}},\ \bibinfo
  {pages} {312} (\bibinfo {year} {2010})}\BibitemShut {NoStop}%
\end{thebibliography}%

\clearpage
\section*{Supplemental information}

\subsection*{Density measurements} 
We derive information about local density changes from the high-angle annular dark-field scanning transmission electron microscopy (HAADF-STEM) signal (electrons collected by the HAADF detector) \cite{Rosner2014}. The dark-field intensity $\frac{I}{I_0}$ contains information about the mass thickness $\rho t$ as follows:
\begin{eqnarray*}
 \frac{I}{I_0} = 1 - \exp \left( - \frac{N_A \sigma \rho  t}{A} \right) = 1 - \exp \left( - \frac{\rho  t}{x_k} \right)
\end{eqnarray*}
and for small arguments:
\begin{eqnarray}
 \frac{I}{I_0} & = & \frac{\rho  t}{x_k}
 \label{eqn:supeq1}
\end{eqnarray}
where $N_A$ is the Avogadro’s number, $\sigma$ is the total scattering cross-section, $\rho$ is the density, $t$ is the foil thickness and $A$ is the atomic weight. $x_k$ is the contrast thickness, which is defined as $\frac{A}{N_A \sigma}$. A simultaneously acquired EEL signal allows calculation of the specimen foil thickness $t$ from the low-loss spectral region \cite{Iakoubovskii2008} and hence the determination of density changes. Using Eq. (\ref{eqn:supeq1}) the relative density change (normalized to the matrix) may be written:
\begin{eqnarray}
 \Delta \rho = \frac{\rho_{SB} - \rho_M}{\rho_M} = \frac{I_{SB}  t_M  x_k^{SB}}{I_M  t_{SB}  x_k^M} - 1
 \label{eqn:supeq2}
\end{eqnarray}
where $\rho_M$, $\rho_{SB}$ are the mass densities, $I_M$, $I_{SB}$ are the HAADF intensities, $x_k^M$, $x_k^{SB}$ are the contrast thicknesses and $t_M$, $t_{SB}$ are the corresponding foil thicknesses for the matrix and the shear band, respectively. The assumption of a constant contrast thickness $x_k$ for matrix and shear band, which cause the $x_k$ terms to cancel in Eq. \ref{eqn:supeq2}, has been discussed in detail elsewhere \cite{Rosner2014}. The calculated foil thicknesses (Fig. \ref{fig:Fig_2}) for the present investigation have been taken as input for calculating the densities in Eq. (\ref{eqn:supeq2}). 

\subsection*{Foil thickness calculation}
The refractive index-corrected Kramers-Kronig sum rule (Eq. (\ref{eqn:supeq3})) was used to calculate the foil thickness based on the low-loss part of each individual EEL spectra \cite{Iakoubovskii2008}:
\begin{eqnarray}
 t = \frac{4 a_0 F E_0}{\left(1 - \frac{1}{n^2} \right) I_{ZLP}} \int \frac{S \left( E \right) \mathrm{d} E }{E \ln \left( 1 + \frac{\beta^2}{\theta_E^2} \right)}
 \label{eqn:supeq3}
\end{eqnarray}
where $S\left(E\right)$ is the single scattering distribution, $I_{ZLP}$ is the integrated intensity of the zero loss peak (ZLP), $n$ is the refractive index of the material ($\sim500$ for metals), $F$ is a relativistic factor, $a_0$ is the Bohr radius, $\beta$ is the collection semiangle, $\theta_E$ is the characteristic scattering angle of inelastic scattering corresponding to an energy loss $E$, and $E_0$ is the microscope voltage in kilovolt units. Calculations using the refractive index-corrected Kramers-Kronig sum rule are performed automatically, for example, within the Digital Micrograph software routines. It analyses the single scattering distribution $S\left(E\right)$, which is obtained from the EEL spectrum by removing the ZLP and plural inelastic scattering using the Fourier-Log method. For data processing the energy-shift of each individual spectrum has been corrected using DM scripts. The thickness at each point has been used in the density calculation (see Eq. (\ref{eqn:supeq2})). 

\subsection*{Zero loss and Plasmon Peak}
\begin{figure}[t]
 a) \includegraphics[width=8cm]{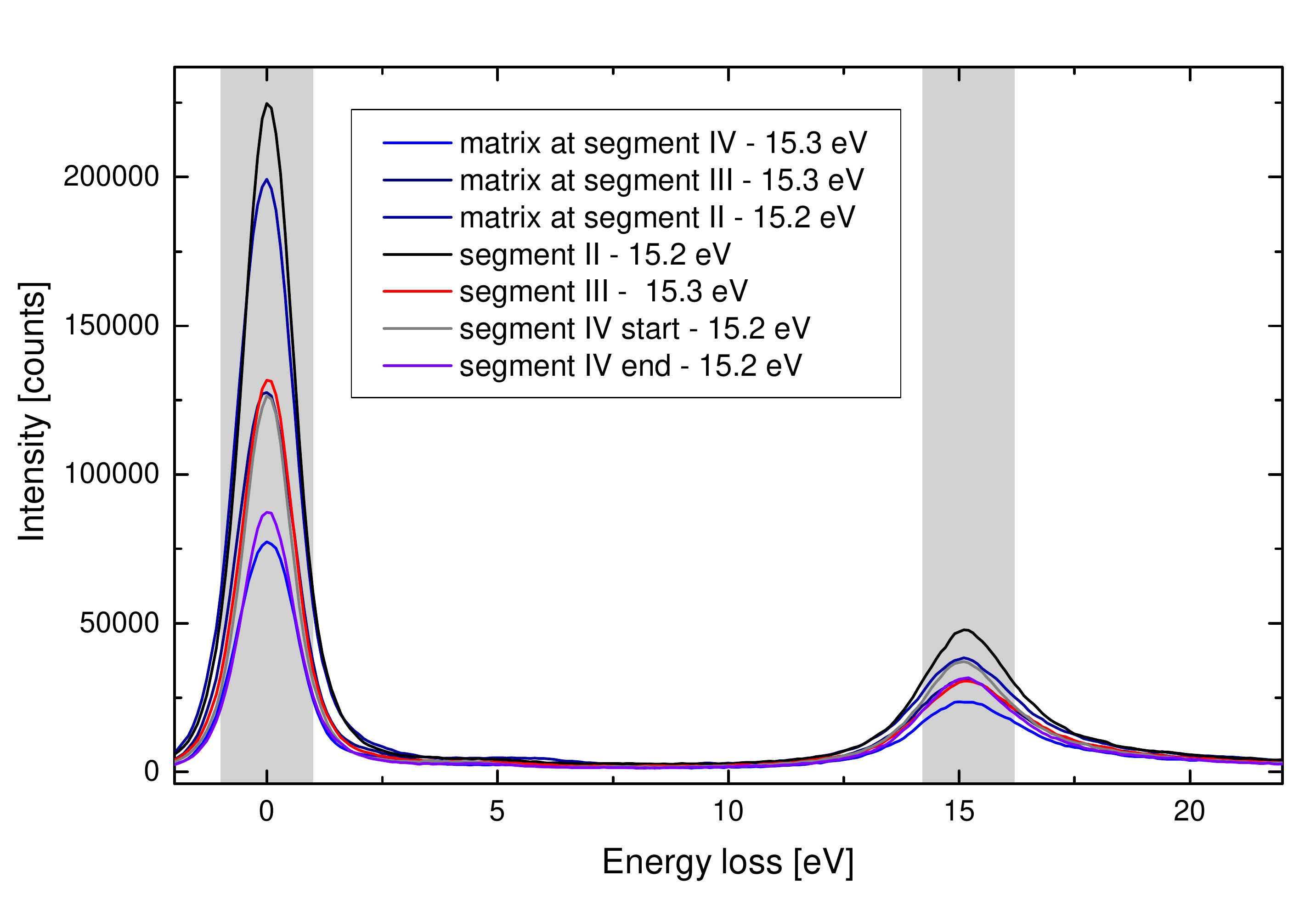} \\
 b) \includegraphics[width=3.8cm]{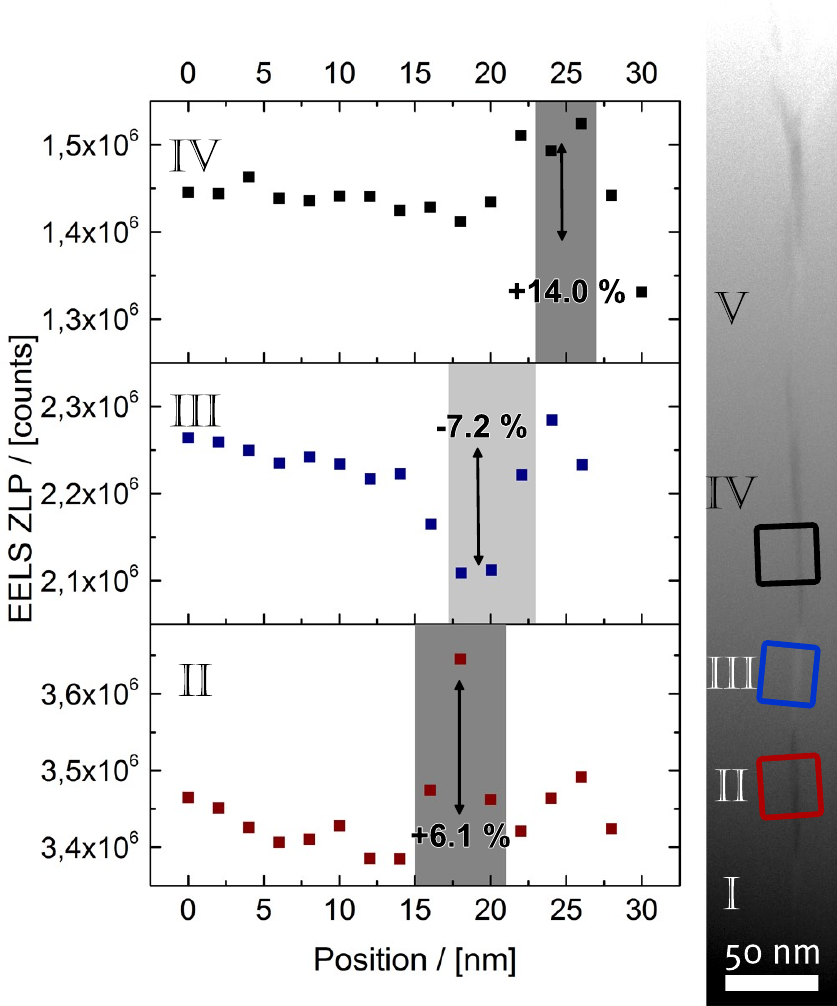} c) \includegraphics[width=3.8cm]{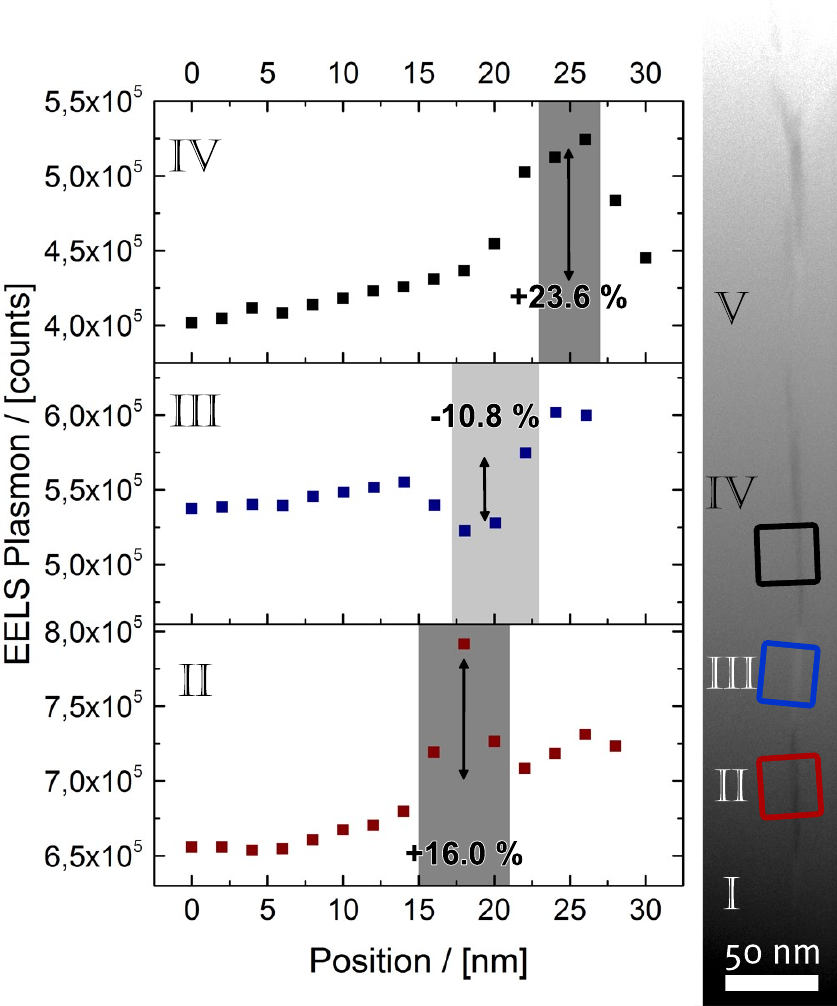} d) \includegraphics[width=3.8cm]{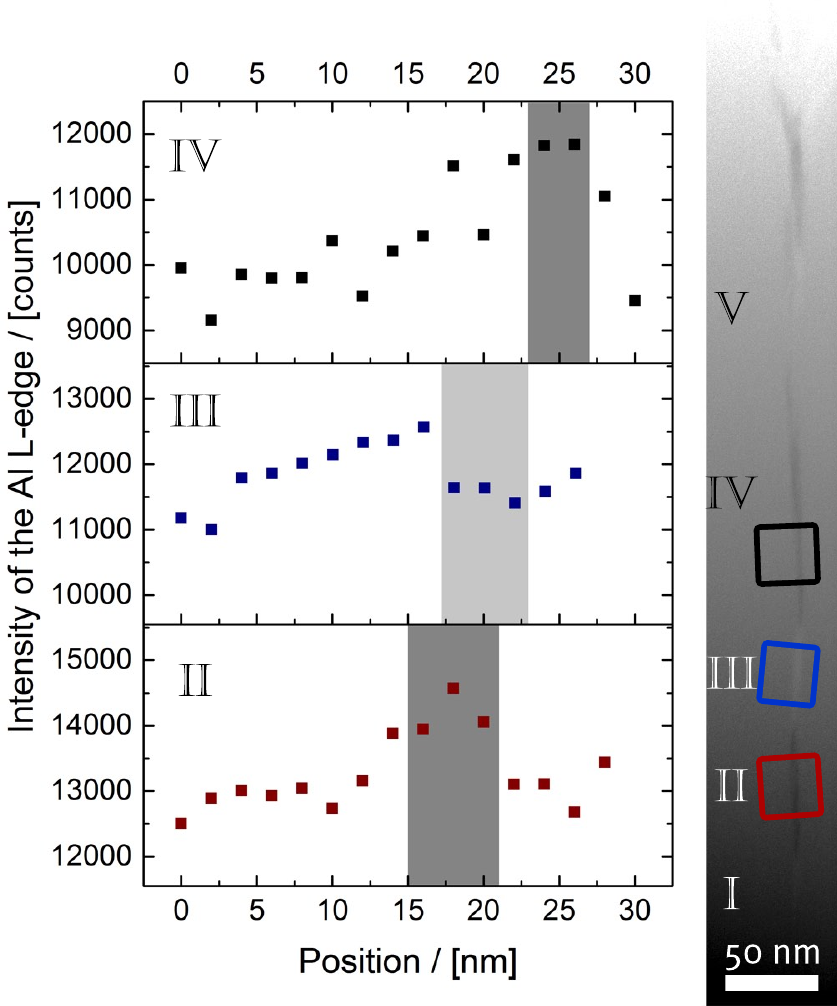}
 \caption{\textbf{Zero loss and plasmon peak/map} (a) EEL spectra for different regions (undeformed matrix, bright/dark shear band segments; see nomenclature in Fig. \ref{fig:Fig_1}). The gray regions indicate the size of the energy window used for extracting the maps. (b), (c) Corresponding Zero loss and plasmon profiles for the different regions. Right: Corresponding regions (boxes) in the HAADF-STEM image. (d) Profiles of the Al-$L$ edge (energy window ranging from 73.5 - 81.5 eV) extracted from the individual EEL spectra. Right: Corresponding regions (boxes) in the HAADF-STEM image.}
 \label{fig:Fig_S1}
\end{figure}

The plasmon peak maximum was found at $15.2 \pm 0.1$ eV by fitting the center of the zero loss peak (ZLP) and the plasmon peak (Lorentzian fits) and subsequently measuring their energy difference. This value confirms results from a FIB-prepared TEM sample \cite{Rosner2014}. An energy shift of the plasmon peak between the shear band segments and the matrix was not observed within the accuracy limit of $\pm0.1$ eV (Fig. \ref{fig:Fig_S1}a). The individual profiles of the Zero loss and plasmon peak are shown in Fig \ref{fig:Fig_S1}b and \ref{fig:Fig_S1}c. Note that the dark shear band regions show an increase in the zero lossintensity which can only be explained by less dense matter. The opposite observation is found for the bright shear band regions which show decreased Zero loss intensities due to denser material. The same statements can be made for the plasmon peaks (Fig. \ref{fig:Fig_S1}c) which is consistent with our previous report \cite{Rosner2014}. A chemical analysis is added here (Fig. \ref{fig:Fig_S1}d) extracted from the Al L-edge using an energy window ranging from 73.5 - 81.5 eV. We find an Al enrichment for the dark shear bands where the pronounced fcc-like structures were observed in the FEM analysis (Fig. \ref{fig:Fig_3}), whereas the bright shear band regions display a slight Al decrease.

\subsection*{Influence of the collection semiangle on the HAADF signal}
\begin{figure}[t]
 \includegraphics[width=8.6cm]{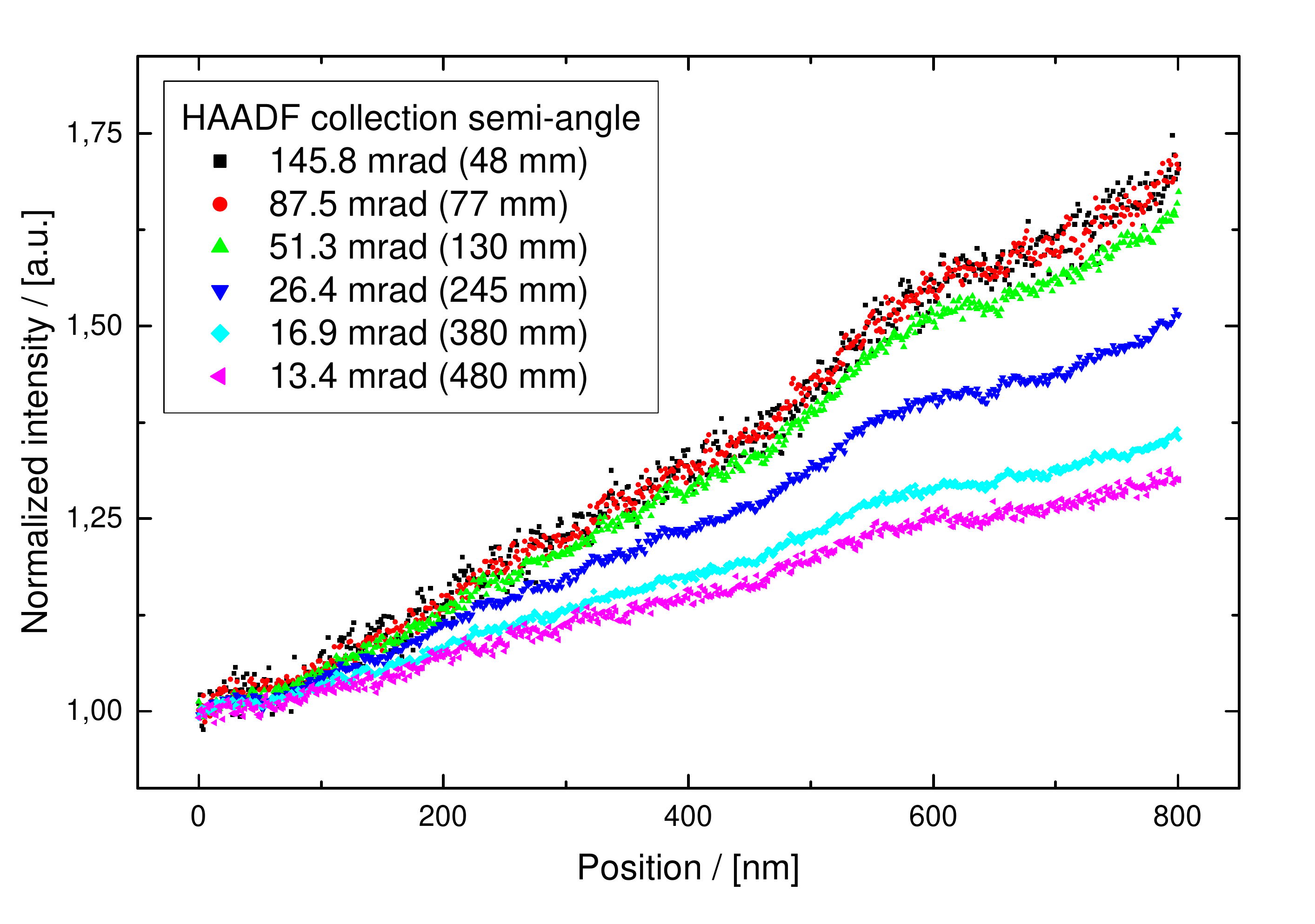}
 \caption{\textbf{Dependence of the HAADF intensity} on the collection semiangle of the HAADF detector. Dark-corrected and normalized HAADF intensity line profiles measured for different collection semiangles of the HAADF detector along a wedge-shaped TEM specimen. The camera lengths are put in parentheses.}
 \label{fig:Fig_S2}
\end{figure}

The collection semiangle affects the collection of electrons contributing to the HAADF signal. This influence has been investigated along the typical wedge-shape of a TEM specimen. The results are shown in Fig. \ref{fig:Fig_S2}. It was found that for collection semiangles larger than 50 mrad, the HAADF signal is essentially constant meaning that the majority of Rutherford-scattered electrons are properly detected. In the present analysis a collection semiangle of 51.3 mrad (130 mm camera length) was used because it offered the best signal to noise ratio. 

\subsection*{Fluctuation electron microscopy (FEM)}
FEM data were obtained from energy-filtered nanobeam diffraction. The intensity fluctuations represented in the diffraction data were analyzed using the annular mean of variance image $\Omega_\textrm{VImage}(k)$ \cite{Daulton2010}. For this purpose, the individual nanobeam diffraction patterns (NBDP) of the region of interest were averaged pixel-by-pixel to a mean and a variance image, which then were converted into annular projections using the PASAD (Profile Analysis of Selected Area Diffraction) tools \cite{Gammer2010} plugin for Digital Micrograph (Gatan). PASAD tools ensure a precise identification of the center of the patterns to perform the annular integration. The final variance $\Omega_\textrm{VImage}(k)$ was calculated by dividing the variance annular average by the mean square. The error was estimated using the standard deviation calculated from several subsets of patterns.

\end{document}